# Direct Observation of Room-Temperature Dislocation Plasticity in Diamond

Anmin Nie[1,5], Yeqiang Bu[2,3,5], Junquan Huang[1], Yecheng Shao[2,3], Yizhi Zhang[2,3], Wentao Hu[1], Jiabin Liu[2,3], Yanbin Wang[4], Bo Xu[1], Zhongyuan Liu[1], Hongtao Wang[2,3]*, Wei Yang[2,3]*, Yongjun Tian[1,6]*

**Affiliations:**

[1]Center for High Pressure Science, State Key Laboratory of Metastable Materials Science and Technology, Yanshan University, Qinhuangdao 066004, China

[2]Center for X-mechanics, Zhejiang University, Hangzhou 310027, China

[3]Institute of Applied Mechanics, Zhejiang University, Hangzhou 310027, China

[4]Center for Advanced Radiation Sources, The University of Chicago, Chicago, IL 60439, USA

[5]These authors contributed equally

[6]Lead Contact

*Correspondence: fhcl@ysu.edu.cn (Yongjun Tian*); yangw@zju.edu.cn (Wei Yang*); htw@zju.edu.cn (Hongtao Wang*)

**Summary**

It is well known that diamond does not deform plastically at room temperature and usually fails in catastrophic brittle fracture. Here we demonstrate room-temperature dislocation plasticity in sub-micrometer sized diamond pillars by *in-situ* mechanical testing in the transmission electron microscope. We document in unprecedented details of spatio-temporal features of the dislocations introduced by the confinement-free compression, including dislocation generation and propagation. Atom-resolved observations with tomographic reconstructions show unequivocally that mixed-type dislocations with Burgers vectors of 1/2<110> are activated in the non-close-packed {001} planes of diamond under uniaxial compression of <111> and <110> directions, respectively, while being activated in the {111} planes under the <100> directional loading, indicating orientation-dependent dislocation plasticity. These results provide new insights into the mechanical behavior of diamond and stimulate reconsideration of the basic deformation mechanism in diamond as well as in other brittle covalent crystals at low temperatures.

**Introduction**

Diamond is the hardest crystalline material [1], with extremely high strength [2,3], widely tunable band-gap [4], and controllable nitrogen-vacancy centers [5]. Such properties of this unique material find a vast range of applications in high-pressure science, machinery industry, electronics and photonics devices, and even biomedicine [6-10]. However, diamond is also the most brittle material due to the strongest C–C covalent bonds [11,12]. As a result, diamond displays virtually no plasticity at room temperature and its brittle nature sets severe limitations in many applications. Understanding the mechanical behavior of diamond, especially its plastic deformation mechanisms at room temperature remains a challenge for decades.

The brittle-versus-ductile response in diamond is attributed to the competition between Griffith cleavage [13] and plastic shear at a crack tip [11,14]. To achieve dislocation slip in diamond, one must first break the strong C–C covalent bond. At room temperature, breaking the C-C bonds tends to lead to cleavage fracture before slip. In addition, preexisting dislocation density in diamond is extremely low, usually several orders of magnitude lower than that in metals. Thus, diamond does not deform with extensive plasticity before cleavage at room temperature [15,16]. The stress state of a material, however, may play an important role in its plastic deformation behavior. Theoretical investigations suggest that high hydrostatic pressure can effectively suppress microcrack propagation in diamond and activate dislocation slip [12,17]. The predicted hydrostatic pressure to trigger the plastic deformation in diamond is as high as several hundred gigapascals [12], which may be realized through indentation [18,19] and compression in the diamond anvil cell [20,21]. Experimentally, dislocations of the {111}<110> slip systems were observed in the vicinity of Knoop indentation [19]. Nevertheless, the possibility that those dislocations were present in the diamond crystal prior to indentation cannot be ruled out [22]. Whether room-temperature plasticity exists in diamond has been debated for several decades due to the lack of direct evidence, which required *in situ* observations [15]. Recent developments of *in situ* mechanical testing techniques in the electron microscope have demonstrated the feasibility of probing elastic deformation [2, 3, 23] and tracking microstructural evolution *in-situ* [24]. In this work, we conducted a comprehensive investigation on plastic deformation of diamond via *in situ* nano-compression experiments in the transmission electron microscope (TEM), combined with atom-resolved TEM observation and three-dimensional (3D) image reconstruction. Extensive dislocation activities in diamond were directly observed in real time. Both {001}<110> and {111}<110> dislocation slip systems can be activated under different loading conditions. Surprisingly, it is easier to activate {001}<110> type slip systems than {111}<110> type, though the latter is more frequently observed in most face-centred-cubic (FCC) crystals [12, 25].

**Results and Discussion**

We performed *in situ* TEM mechanical tests at room temperature on sub-micron-sized diamond pillars, which were prepared from Type Ib diamond single crystals (Henan Famous Diamond Industrial Co.) with focus ion beam (FIB) milling followed by Argon plasma thinning (fig. S1A-D). Figure 1A shows a as-prepared <111>-oriented pillar with a thinner head over a thicker and cylindrical body (see fig. S1E as well). The preparation of pillars with <110> and <100> orientations follows the same procedure. A layer of amorphous carbon was present over the as-prepared pillars (Fig. 1B) due to FIB-milling. The thickness of the layer was less than 2 nm, with essentially no influence on the mechanical properties of the pillar. The cylindrical axis of the pillars was along the <111> direction, as confirmed by selected-area electron diffraction (SAED) (inset to Fig. 1B). High-angle annular dark-field STEM (HAADF-STEM) image (Fig. 1C) shows nearly perfect atomic arrangements in the diamond pillar prior to the test. During compression, the thinner head of the pillar was partially broken off from the thicker shoulder due to stress concentration (movie S1), leaving a fresh and sharp $[\bar{1}\bar{1}1]$ fracture surface over the pillar top, as indicated by the weak beam dark-field TEM image (Fig. 1D). Further compression was then loaded on the freshly fractured surface (movie S2). Several dislocation half-loops were first developed from the fracture surface (Fig. 1E). Driven by the compressive loading, these half-loops then multiplicated and propagated from the tip into the lower part of the pillar (Fig. 1F). Individual dislocation half-loops can be divided into two types of segments (inset of Fig. 1F), which are referred to as head and arm segments, respectively. The generated dislocations propagated in several gliding planes. The two arm segments slipped in planes parallel to the sideface of the pillar with a speed faster than that of the head segment, which slipped toward the bottom of the pillar. With increasing load, dislocation density increased significantly; multiple dislocation sources were activated, indicating that the fresh fracture surface was an effective germination site for dislocations (Fig. 1G). Similar observations were repeated in several other diamond pillars tested (movies S3 and S4, figs. S2 and S3), demonstrating a universality of the dislocation generation in diamond at room temperature during compression along the <111> direction.

Each compressed diamond pillar was tilted into various two-beam conditions, in order to determine Burgers vectors of the dislocations (Fig. 2). For the pillar shown in Fig. 1, multiple dislocation arrays (α, β, and γ in Fig. 2) in different slip planes were clearly distinguished. Three images were taken under different two-beam conditions, with $\boldsymbol{g}=[\bar{1}11]$ along the [101] zone axis (Fig. 2A), and $\boldsymbol{g}=[\bar{1}\bar{1}1]$ (Fig. 2B) and $\boldsymbol{g}=[\bar{2}20]$ (Fig. 2C) along the [112] zone axis. Dislocation array α remained visible under all two-beam conditions. In contrast, array β was invisible under $\boldsymbol{g}=[\bar{1}11]$, whereas array γ was only visible under $\boldsymbol{g}=[\bar{1}\bar{1}1]$. According to the extinction criterion $\mathbf{g} \cdot \mathbf{b} = 0$ (where $\mathbf{b}$ is the Burgers vector), the Burgers vectors were determined to be 1/2$[\bar{1}01]$, 1/2 $[01\bar{1}]$, and 1/2[110] for arrays α, β, and γ, respectively. The same 1/2<110> Burgers vectors were also confirmed in other diamond pillars tested (figs. S4 and S5). Further tilting the diamond pillar to the [101] zone axis led to the head segments of dislocation array α nearly parallel to the incident beam (fig. S6), under which condition atom-resolved HAADF-STEM images were taken to investigate the dislocation core structures. The closure failure of the Burgers circuit reveals a Burgers vector of 1/2 $[\bar{1}01]$ for the head segments of dislocation array α (Fig. 2D), further confirming the aforementioned $\mathbf{g} \cdot \mathbf{b}$ analysis. In addition, the core structure of the head segments matches well with those of the edge dislocations in diamond lattice [26], rather than the common screw or 60° mixed-type dislocations.

To clarify slip planes of the generated dislocations and configurations of the arrays, we rotated the diamond pillar around the [111] axis from 0 to 180°, and recorded TEM images at various rotation angles (Fig. 3). Dislocation arrays α, β, and γ were all clearly visible at a rotation angle of 26° (Fig. 3A). At 47°, the slip plane of dislocation array β was nearly edge-on and parallel to the (100) plane according to the corresponding SAED pattern (Fig. 3B). Increasing rotation angle further revealed that slip planes of arrays γ (Figs. 3C) and α (Fig. 3D) were parallel to the (001) and (010) planes, respectively. Combining the slip plane determination based on 3D tomography (as illustrated in Figs. 3E–H) with Burgers vectors information from two-beam diffraction and HAADF-STEM (Fig. 2), it is clear that the {001}<110> slip systems were activated under compression in the <111>-oriented diamond pillars in our tests, with the Burgers vectors (yellow arrows in Fig. 3B–D) lying in the slip planes. The coexistence of curvilinear dislocation loops and in-plane Burgers vector suggests that the dislocation lines contain components both parallel (screw) and perpendicular (edge) to the Burgers vectors. This analysis is also consistent with the atom-resolved TEM observation (Fig. 2D), where the screw-type arm segments were difficult to image at atomic level due to the non-planar core structure. Similar slip-trace analyses (figs. S7 and S8) and the 3D tomography of dislocations (movie S5) of other pillars further confirmed these results.

We further investigate the dislocation behaviors in the diamond nanopillars with other orientations, including <110> and <100>, respectively. As shown in Fig. 4A, the dislocations are also generated in the <110> oriented diamond nanopillar under uniaxial compression. The generated dislocations are determined to be in the (010) plane after a 40° rotation of the nanopillar. Interestingly, despite the Schmid factor of {100}<110> slip system is lower than that of {111}<110> slip system under the <110> direction loading (Table S1), the dislocations still slip in {100} plane. Fig.4 B shows the uniaxial compression of the <100> oriented diamond nanopillar, in which the dislocations in the {111} planes are activated. In the case of <100> compression, the Schmid factor of {100}<011> slip system is 0, possibly answering the domination of the {111} plane slips in the <100>-compressed diamond with no occurring of the {100} plane slip. On the other hand, the in-situ compression of <111> orientated silicon nano-pillar that possesses the same lattice structure as diamond was also investigated in our experiments. As shown in Fig. S10, the tomographic reconstruction indicates the dislocations activate in the {111} plane, differing from that of diamond but consistent with the observation in typical FCC metals[27]. Above all, the unusual dislocation behaviors in diamond are not only related to the Schmid factor, but also determined by its intrinsic nature, *e.g.* lattice parameter and strong C-C covalent bonds *etc.*. Though more theoretical research is needed in the future studies, our experimental investigation will trigger a rethink of the plastic deformation mechanism of diamonds.

Thus, our experimental investigations indicate that 1/2<110>{001} dislocations dominate deformation in diamond at room temperature under uniaxial compression of both <111>- and

<110>-oriented diamond pillars. The <110>{111} slip system is only activated in the case that the resolved shear stress on {100} planes is zero. On one hand, both *in situ* mechanical experiments [2, 3, 23] and theoretical studies [16, 20] indicate that diamond possesses the highest brittleness and exhibits no plastic deformation, even in nanopillars under similar uniaxial loading [23]. On the other hand, the close-packed {111} planes are usually deemed to be the prime slip plane in FCC crystal at room temperature[13, 26]. It has been experimentally observed the plasticity dominated by the {111}<110> slips in the compression of nickel nanopillars with the <111> orientation[27]. The unusual {100} plane slip was only occasionally observed at elevate temperature for some FCC crystals, *e.g.* silicon[28] and aluminum[29]. However, dislocation plasticity dominated by {100}<110> slip system has been rarely observed at room temperature in FCC crystals, especially in the strong covalent crystals. The activation of such unusual slip is attributed to higher Schmid factor of {100}<110> slip system, higher stress can be approached in small-sized crystals and more thermal activation effect from higher temperature[28]. To understand the origin of plasticity observed in the tested pillars, we examine elastic instabilities for the {001}<110> and {111}<110> slip systems as well as {111} cleavage, using a criterion based on free energy consideration [30, 31], which also takes Schmid factor and stress state into consideration. To derive the criterion of elastic instability before stress relaxation occurs in a given system, we apply a plane wave perturbation to the second-order derivative of free energy of a volume element under a given displacement of deformation, following [30, 31]

$$\Lambda(\boldsymbol{\omega},\boldsymbol{n}) \equiv (C_{ijkl}\omega_i\omega_k + \tau_{jl})n_j n_l , \tag{1}$$

where $\Lambda(\boldsymbol{\omega}, \boldsymbol{n})$ represents second-order derivative of the free energy for a plane-wave displacement perturbation in the Cartesian coordinate $\boldsymbol{x}$ in the form of $\boldsymbol{\omega} \exp(i\boldsymbol{n}\cdot\boldsymbol{x})$, where $\boldsymbol{\omega}$ is a unit vector representing the direction of the perturbating and $\boldsymbol{n}$ is the wavevector [31], $C$ is the fourth-order elastic constant tensor at a given deformation state, and $\tau$ is the internal (Cauchy) stress tensor. We note that $\Lambda(\boldsymbol{\omega}, \boldsymbol{n})$ is essentially a more elaborate Schimid factor that considers effect from stress, geometry and instantaneous elastic constants. The condition for maintaining elastic stability is $\Lambda(\boldsymbol{\omega}, \boldsymbol{n}) > 0$. Once $\Lambda(\boldsymbol{\omega}, \boldsymbol{n})$ drops below zero for a given pair of $\boldsymbol{\omega}$ and $\boldsymbol{n}$, the lattice reaches instability and defects will nucleate, leading to stress relaxation. The possible modes of stress relaxation depend on the configuration of the ($\boldsymbol{\omega}$, $\boldsymbol{n}$) pair: Brittle cleavage occurs in the plane with normal $\boldsymbol{n}$ when $\boldsymbol{\omega}$ is parallel to $\boldsymbol{n}$; a dislocation slip system along $\boldsymbol{\omega}$ is activated in the plane with normal $\boldsymbol{n}$ when $\boldsymbol{\omega}$ is perpendicular to $\boldsymbol{n}$.

We calculate elastic constants and stress tensors of the diamond lattice with a series of compressive strains using first principles. The first stress drops in the stress-strain curve with compressive load in the [111] direction of diamond lattice occurs at a strain of 27% (Fig. 5A), indicating onset of elastic instability and initiation of defects (dislocations or microcracks) at that stress level. We then calculate $\Lambda(\boldsymbol{\omega}, \boldsymbol{n})$ as a function of strain with three corresponding ($\boldsymbol{\omega}$, $\boldsymbol{n}$) pairs for the (100)[011] and $(11\bar{1})[011]$ slip systems and the $(11\bar{1})$ cleavage. $\Lambda(\boldsymbol{\omega}, \boldsymbol{n})$ of the (100)[011] slip system is the first to become negative at a strain of 27% (Fig. 5B), corresponding to the first stress drop of the stress-strain curves in Fig. 5A. This indicates that (100)[011] slip system is indeed the dominant stress relaxation mode prior to $(11\bar{1})[011]$ slip system and $(11\bar{1})$ cleavage. Once this relaxation mode is activated, other modes are not needed for further deformation in single-crystal pillars. Li et. al.[30-32] proposed a "B criteria" to predict the lattice stability under arbitrary but uniform external load. In our case of diamond compressed along <111> direction, the "B criteria" could be expressed as:

$$B_{66} = C_{66} + 1/2\ (\sigma_{11}+\sigma_{22}) \qquad (2)$$

where $B_{66}$ is the component of elastic stiffness coefficients tensor, $C_{66}$ is the component of elastic tensor, which is equal to shear modulus ($G$), $\sigma_{11}$ and $\sigma_{12}$ are two components of stress tensor. The lattice is elastic instability under such loading conditions as soon as $B_{66} = 0$, i.e.,

$$C_{66} = -1/2\ (\sigma_{11}+\sigma_{22}) \tag{3}$$

Diamond lattice is severely deformed at the critical strain (~27%) of elasticity. According to our first principle calculations, the loading direction (i.e., <111>) has an angle of ~45° to both the slip plane (i.e., (100) plane) and slip direction (i.e., <110>) in such severely deformed lattice (as shown in Fig. S9). According to the Schmid law, at the critical strain, the shear stress ($\tau$) applied on the {100}<110> slip system approach the maximum value, i.e.,

$$\tau \approx -1/2\ \sigma \tag{4}$$

where $\sigma$ is the applied normal stress. As $\sigma_{22} = 0$ for confinement-free compression, we can derive that the lattice is unstable when $G \approx \tau$ for this case. Figure 5C shows the shear modulus $G_{[011](100)}$ and resolved shear stress $\tau_{[011](100)}$ corresponding to the {100}<110> slip in the deformed crystal lattice coordinate under [111] direction compression. It is noted that $G_{[011](100)}$ is equal to $\tau_{[011](100)}$ at the critical strain of 27%. This again prove that the occurrence of the elastic instability of this perturbation mode.

The dislocations in {111} slip planes of diamond have been theoretically investigated in early study[33, 34], showing the shuffle-set and glide-set. Here, we further investigate the atomic mechanism of the {100}<110> slip process by density functional theory based molecular dynamics simulations. Figure 6A shows a preset dislocation centered the simulation box with Burgers vector 1/2[110] lying in the [001] plane. The core structure is built according to our experimental observation (Fig. 2D). The Cartesian coordinate system is set as x = [110], y = [001] and z = [-110]. The external loading is given in the form $\sigma_{ij} = -p\ \delta_{ij} + \tau_{12}(\delta_{i1}\ \delta_{j2} + \delta_{i2}\ \delta_{j1})$, where $p$ is the hydrostatic pressure, $\tau_{12}$ the shear stress and $\delta_{ij}$ the Kronecker delta. For each simulation, the hydrostatic pressure is fixed while shear stress is increased in a stepwise manner. Given the state of pure shear, *i.e.* $p = 0$, the dislocation remains still until C-C bond breakage is initiated at the dislocation core at $\gamma_{12} = 3.7\%$ (Fig. 6B). Further increasing the load leads to (111) cleavage, which is clearly caused by the positive resolved normal stress on the (111) plane. To suppress the positive normal stress component on all atomic planes, a large hydrostatic pressure is thus applied. Given $p = 400$ GPa, no crack can be nucleated and the dislocation slip at $\gamma_{12} = 11.9\%$ (Fig. 6C1 and C2). Figure 6D elucidates the mechanism of dislocation slip by overlapping two configurations before and after advancing the dislocation by one Burgers vector. The atomic displacement is featured by the rotation of the C-C bond in the dislocation core, as indicated by arrows in Fig. 6D. The associated bond breakage and rebonding process relocates the five- and seven-membered rings to neighboring sites, and realizes dislocation slip by 1/2[110]. Sequential bond rotation leads to continuous dislocation motion.

**Conclusion**

In summary, we directly observed room-temperature plasticity in diamond single crystals under confinement-free compressive deformation, with unambiguous information of the type, structure, and motion of the generated dislocations. The identified plasticity in diamond is dominated by dislocations slipping in the non-close-packed {100} planes under uniaxial compression of <111> and <110> directions, respectively. Such slip systems have rarely been recognized or considered for FCC crystals at room temperature. Moreover, typical dislocations in {111} planes are generated in the uniaxial compression on <100> orientated one, indicating orientation dependent dislocation behaviors in diamond. Our new results and the technique developed in current work can be extended to understand the deformation behavior of other brittle covalent crystals.

**Experimental Procedures**

**Sample preparation.** The starting material was a Type Ib diamond monocrystal from Henan Famous Diamond Industrial Co., produced by high-pressure, high-temperature synthesis. A diamond sheet (10×5×2 μm$^3$) was cut from the (111) facet of the diamond monocrystal with FIB milling (3.0 nA, 30 kV), and then transported onto a TEM half grid with lift-out technique for further processing into step-like sub-micron-sized pillars by FIB milling (0.1 nA, 10 kV). Typical SEM images during the sample preparation are displayed in Fig. S1A–D. Residual amorphous carbon and irradiation layer over as-milled pillars were reduced by sequential argon plasma thinning with gradually decreasing voltage from 1.0 kV to 0.8 kV for 4 hours.

**TEM characterization.** *In situ* uniaxial compression experiments were carried out in a JEM 2100 microscope equipped with a homemade X-Nano TEM mechanical stage[3, 35]. The sample was mounted on the nano-manipulator end of the X-Nano stage, and precisely driven against a diamond indenter in a stepwise mode with built-in piezo actuators (positioning accuracy ~ 0.1 nm). The loading setup and operation of X-Nano TEM stage are schematically drawn in Fig. S1F. The X-Nano TEM mechanical stage possesses four-degrees of freedom, *i.e.*, 360-degree rotation and three-dimensional positioning of the sample. This allowed us to observe spatial configurations of dislocations and determine slip planes. Atomic-level characterization was performed with an aberration-corrected FEI Themis Z STEM.

**The first principle calculations.** The Vienna ab initio simulation package (VASP) was employed to perform first principles calculations, based on density functional theory within the plane-wave pseudopotential approach[36]. The cutoff energy for the plane wave was 400 eV, and an 8×8×8

Monkhorst-Pack K-point mesh was used in Brillouin zone sampling. The appropriate cuboid unit cells used in strain-stress calculations were relaxed to reduce undesired stress component to less than 0.02 GPa.

**The density functional theory based molecular dynamics simulations.** The density functional based tight binding (DFTB) method that is implemented in the DFTB+ package[37] was used for molecular dynamics simulations. The DFTB parameters used in the simulations are optimized for periodic boundary conditions[38]. A system containing 794 carbon atoms under periodic boundary condition was used, and an edge dislocation dipole was initially placed and relaxed. Quasi-static loading with shear strain was applied to the system. In each loading step, a shear strain of approximately 0.2% was applied to the systems, following a relaxation. During the whole simulation, self consistent charge (SCC) calculation was carried out and an $1\times4\times1$ Monkhorst-Pack K-point mesh was used in Brillouin zone sampling. All relaxation process reduced the max force component to less than 0.02 meV/Å.

**Supplementary Materials**
Supplemental Information can be found online.

**Acknowledgments:**
This work was supported by the National Key R&D Program of China (2018YFA0703400), and the National Natural Science Foundation of China (11725210, 1170216, 11672355 and 51672239).

**Author contributions:**
Y.J.T. and W.Y. initiated the project and created the experimental protocols. H.T.W., Y.Z.Z and J.B.L developed XNano in-situ TEM holder. J.Q.H., Y.Q.B. and A.M.N. carried out the fabrication of diamond nanopillars. Y.Q.B. and A.M.N. conducted the *in-situ* TEM testing. Y.C.S., Y.Q.B. and H.T.W. performed DFT calculations. Z.Y.L, B.X., W.T.W and Y.B.W analyzed the data.

A.M.N., H.T.W., Y.J.T. and W.Y. wrote the manuscript and all the authors contributed to the discussion and revision of the manuscript.

**Declaration of Interests:**

The authors declare no competing interests.

**Data Statement:**

All data are reported in the paper or the supplementary materials

**Figure Legends**

**Figure 1. Evolution of a diamond nanopillar during compression.** (A) Bright-field (BF) TEM image of a diamond nanopillar before compression. (B and C) are atomic-scale BF and HAADF STEM images of the diamond nanopillar, respectively. Inset of (B) shows a SAED pattern of the diamond nanopillar. Inset of C shows enlarged atomic-scale HAADF image of the diamond nanopillar. (D-F) Weak beam dark-field (DF) TEM images showing dislocation evolution in the diamond nanopillar during compression. Inset of panel (F) illustrates schematically the half-loops. Head and arm segments are colored by red (I) and orange (II), respectively.

**Figure 2. The determination of the Burgers vectors of dislocations.** (A) was taken under $g = [\bar{1}11]$ two-beam condition along the [101] zone axis; (B and C) were taken under $g = [\bar{1}\bar{1}1]$ and $g = [\bar{2}20]$ two-beam condition along the [112] zone axis, respectively; (D) Atomically resolved HAADF-STEM image showing a dislocation core in after in situ deformation.

**Figure 3. Three dimensional configurations of the generated dislocations and their slip planes in diamond nanopillar.** (A-D) DF-TEM images of the compressed diamond nanopillar tilted at various angles. (E-H) are schematic drawings for the dislocations and their gliding planes as observed in (A-D).

**Figure 4. Dislocation behaviors in diamond under various loading directions.** (A1 and A2) Dislocations are activated in diamond under loading with <110> direction. (A3) The slip plane is rotated to its edge-on view, and determined to be (010) plane. (B1 and B2) The dislocations are activated in diamond under loading with <100> direction. (B3) The slip plane is rotated to its edge-on view, and determined to be (111) plane.

**Figure 5. Elastic stability criteria for determining the mode of incipient plasticity.** (A) Theoretical stress-strain curve of uniaxial compression along [111] for a diamond lattice, the first stress drop appears at strain of 27%; (B) $\Lambda(\omega, n)$ for the three perturbation modes, *i.e.* (100)[011] slip system, $(11\bar{1})[011]$ slip system and $(11\bar{1})$ cleavage. (C) Elastic moduli and resolved stresses corresponding to the three relaxation modes. The subscripts indicate the corresponding $(\omega, n)$ pairs.

**Fig. 6. The density functional theory based molecular dynamics simulations for the diamond under different loading conditions.** (A) The model of diamond lattice containing an edge dislocation with Burgers' vector 1/2[110] in the (001) plane after relaxation. The dislocation core is highlighted. (B) Crack initiates under a pure shear strain of 3.7% as indicated by C-C bond breakage at the edge dislocation core. The defected region is highlighted. (C1 and C2) edge dislocation slips in [001] plane by one Burgers' vector driven by a shear strain of 11.7% under 400GPa hydrostatic pressure. The red and yellow highlight the dislocation core locations before and after slipping forward in one Burgers vector, as indicated by the arrow. (D) The C1 and C2 are overlapped to show the mechanism of dislocation slip. The dislocation cores are highlighted in red and yellow, respectively. The two arrows indicate the major position change of C atoms, which is characterized as a bond rotation process.

**Supplemental Movies**

**Movie S1.** The top part of the diamond nanopillar was broken during compression (Fig. 1D is from this video). Video speed at 3 times the speed of experiment.

**Movie S2.** The dislocations slipping on multiple planes were activated from the fracture surface during the further compression (Figs. 1E-G are from this video). Video speed at 3 times the speed of experiment.

**Movie S3.** The second example showing the multiplication and movement of dislocations in diamond (fig. S1 is from this video). Video speed at 3 times the speed of experiment.

**Movie S4.** The third example showing the multiplication and movement of dislocations in diamond (fig. S2 is from this video). Video speed at 3 times the speed of experiment.

**Movie S5.** The spatial configuration of dislocations in diamond (fig. S6 is from this video). The sample was rotated from 0º to 180º at 1º increment around the [111] axis.

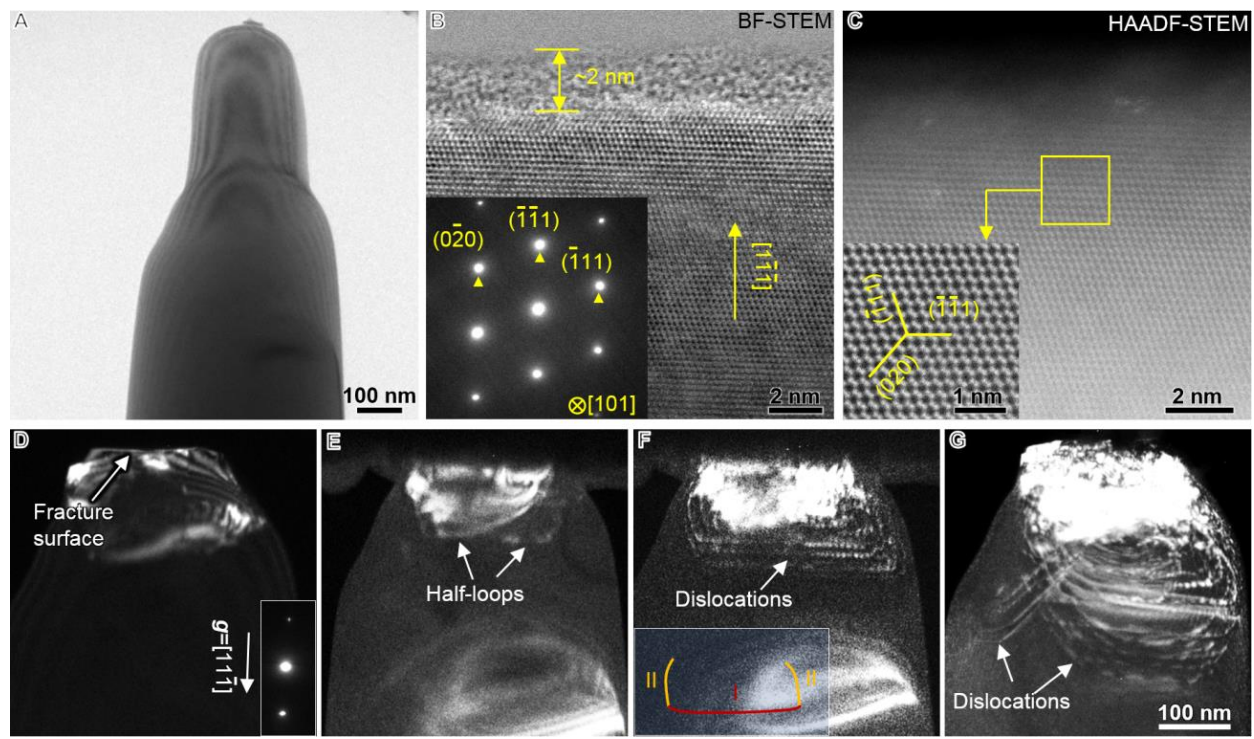

**Fig. 1.**

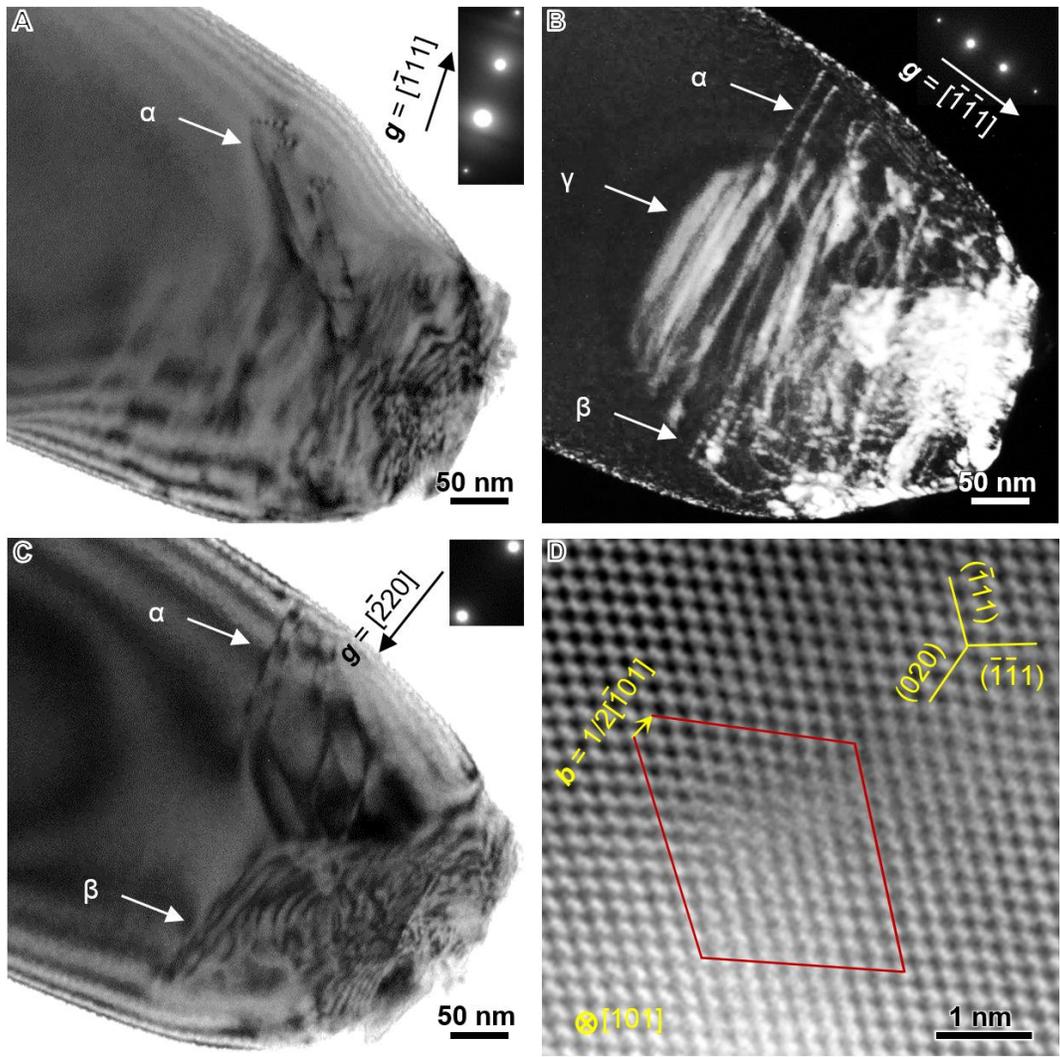

**Fig. 2.**

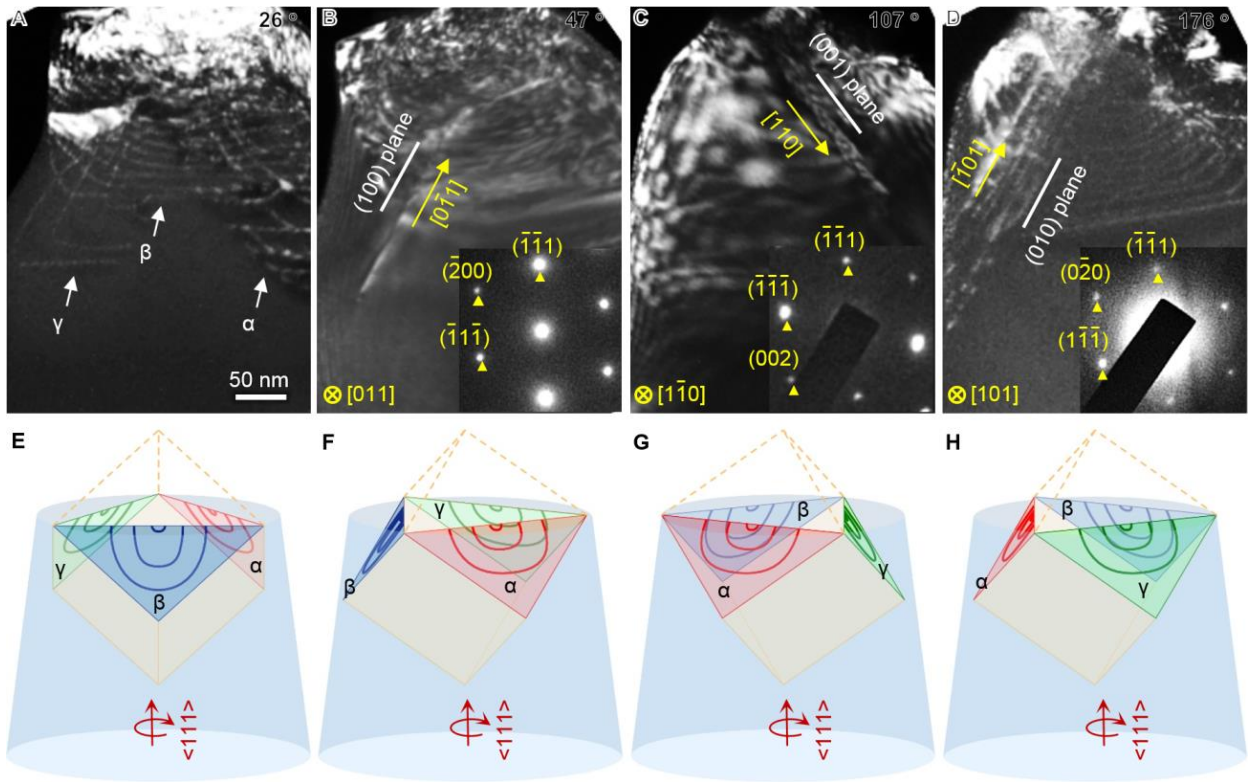

**Fig. 3.**

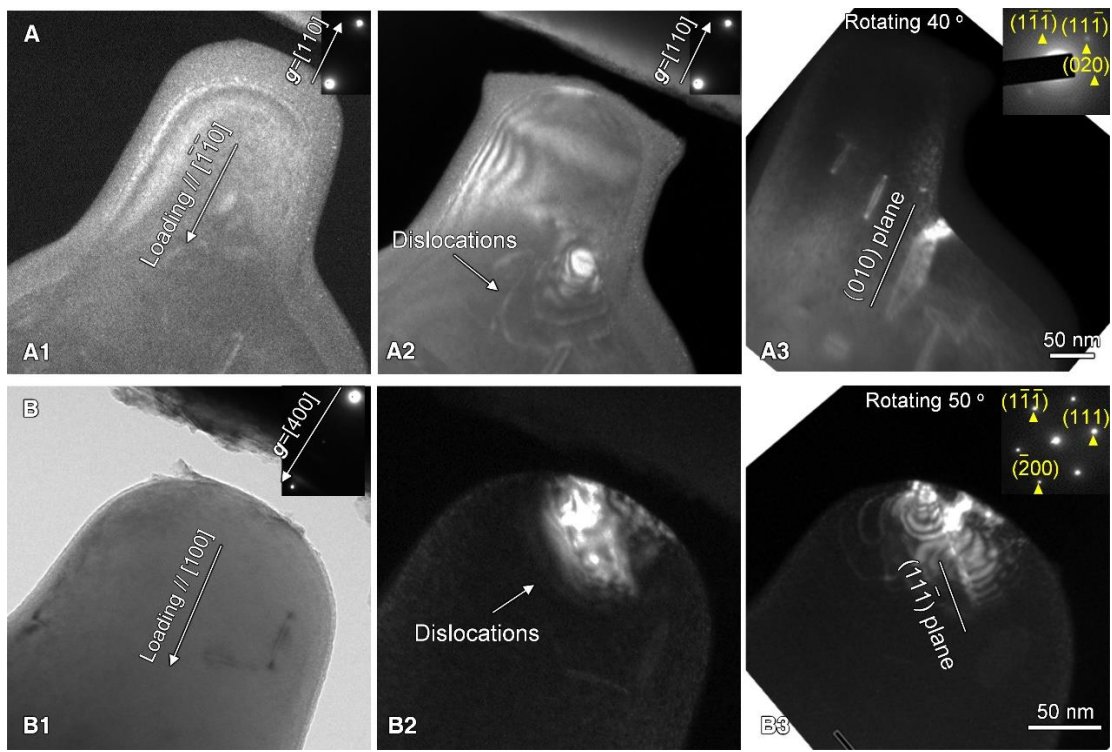

**Fig. 4.**

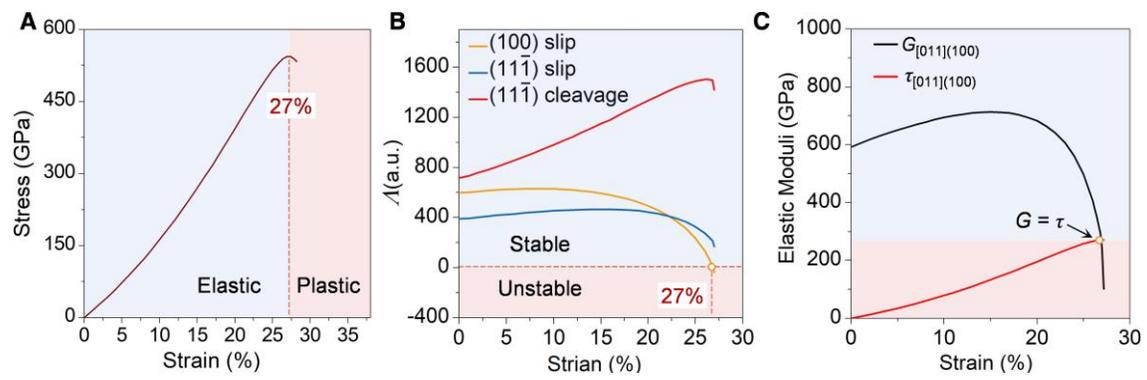

**Fig. 5.**

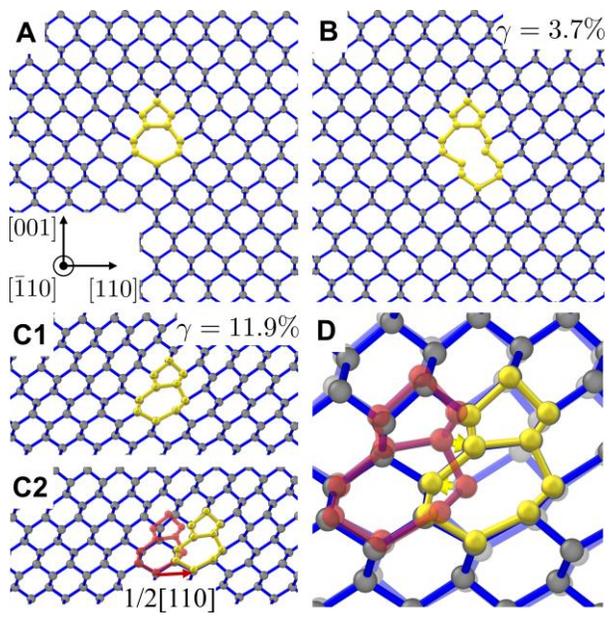

**Fig. 6**

**SUPPLEMENTAL DATA**

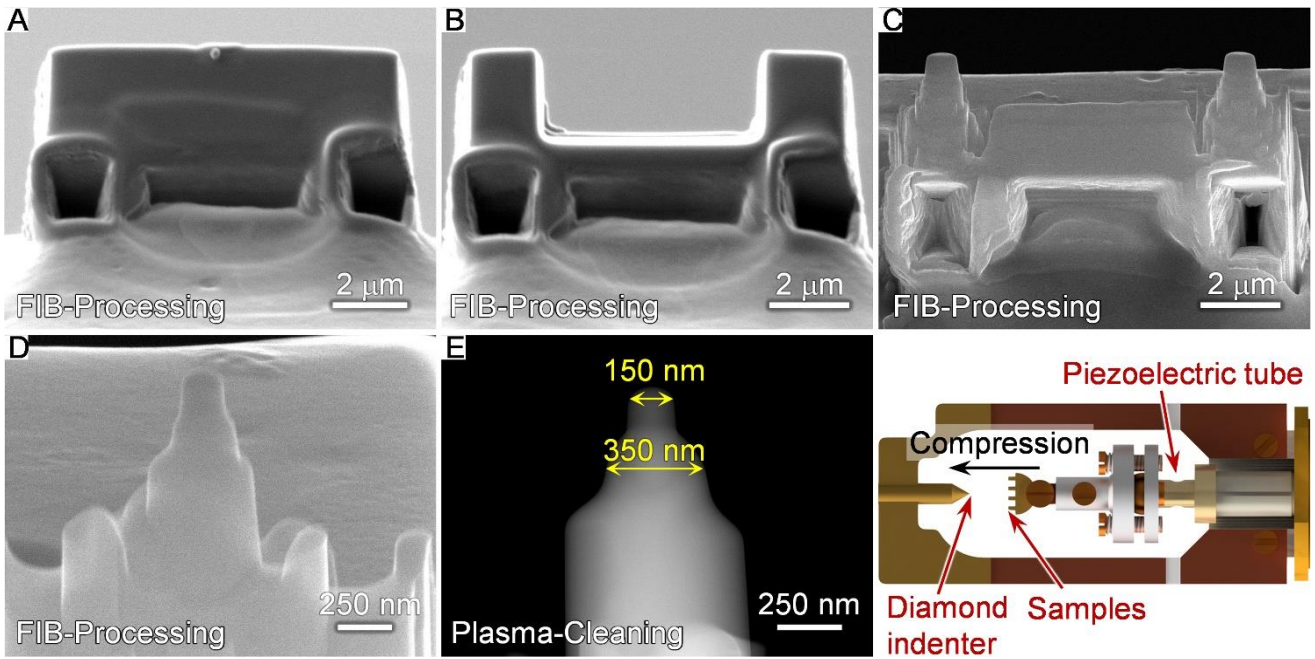

**Figure S1. Diamond pillar fabrication and loading setup.** (A) A diamond sheet extracted from a diamond monocrystal with FIB milling. (B-D) Snapshots during the FIB-processing to fabricate step-like diamond pillars. (E) HAADF-STEM image showing a typical diamond pillar after plasma cleaning that is ready for in situ tests. (F) Schematic drawing of the loading setup and the operation of X-Nano TEM mechanical stage.

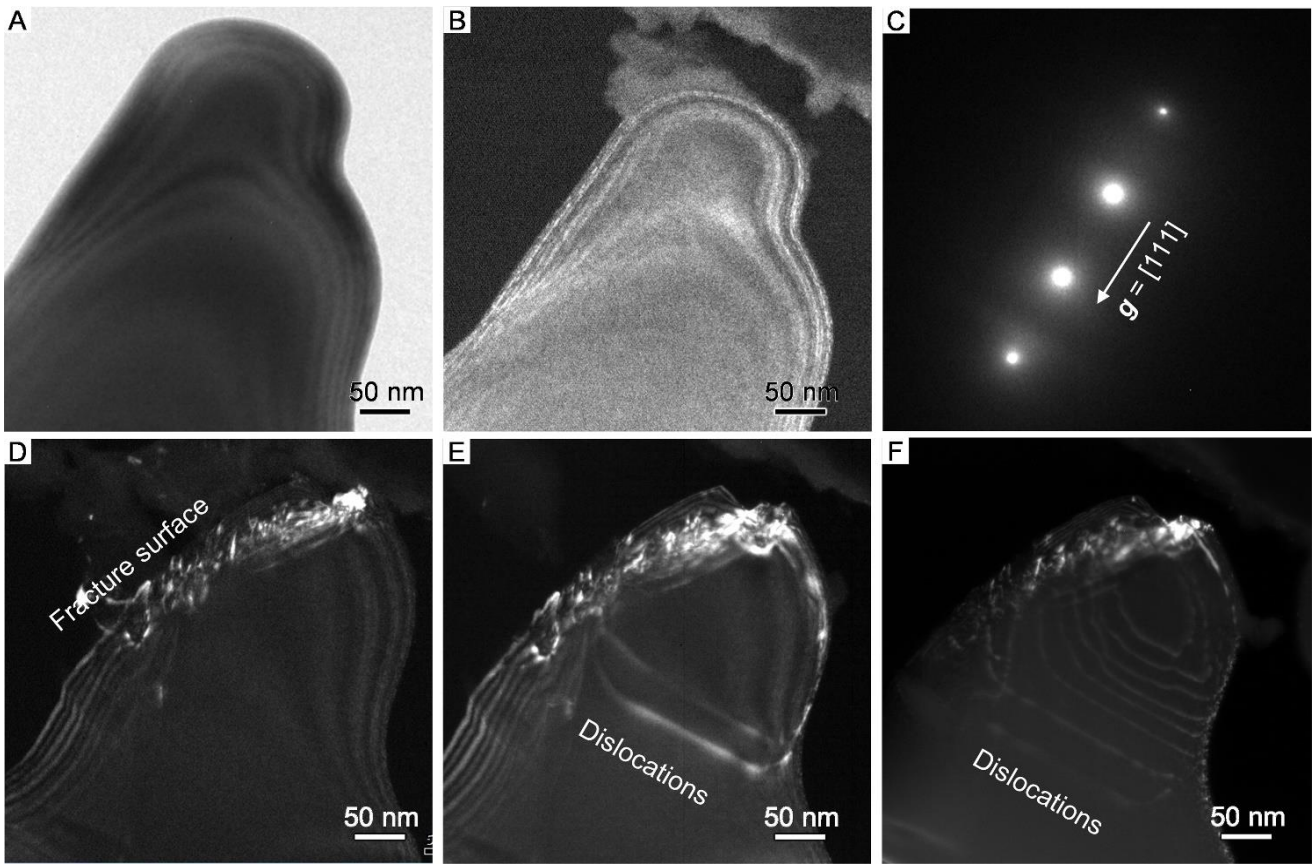

**Figure S2. Multiplication and motion of dislocations in a diamond pillar.** (A and B) Bright-field (BF) and dark-field (DF) TEM images of the diamond pillar. (C) The corresponding selected area electron diffraction (SAED) pattern, showing that the [111] axis is along the axis of this diamond pillar. The *in situ* process was captured under ***g*** = [111] two-beam condition. (D) A fracture surface of diamond due to first-stage compression. (E and F) Dislocations emitted from the fracture surface in subsequent deformation.

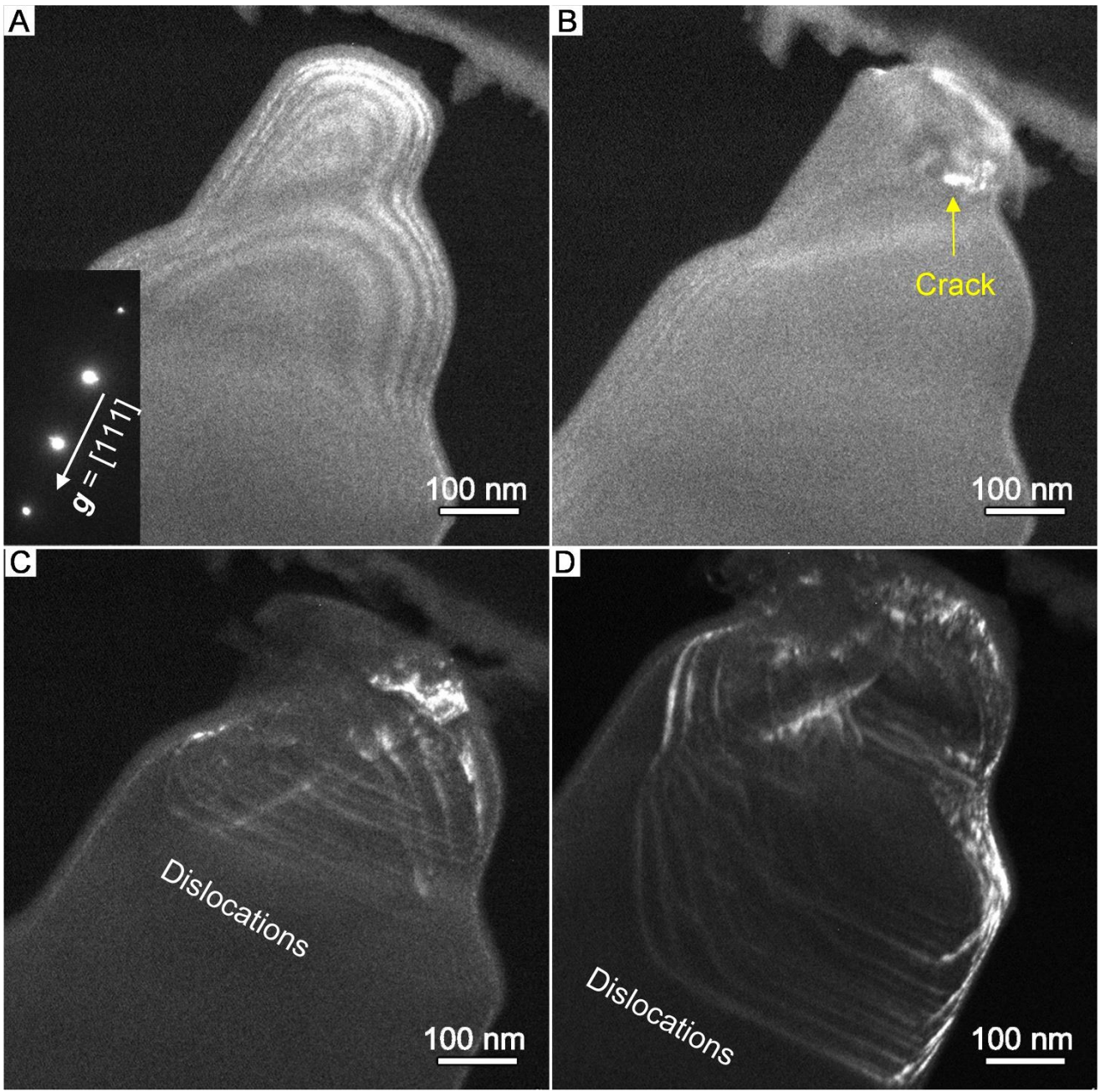

**Figure S3. Multiplication and motion of dislocations in a diamond pillar.** (A) The DF-TEM image of a [111]-oriented diamond pillar, inset shows the corresponding SAED. (B) A crack initiated from the contact point. (C) Dislocations slipping on multiple planes are activated near the crack. (D) Multiplication and motion of dislocations.

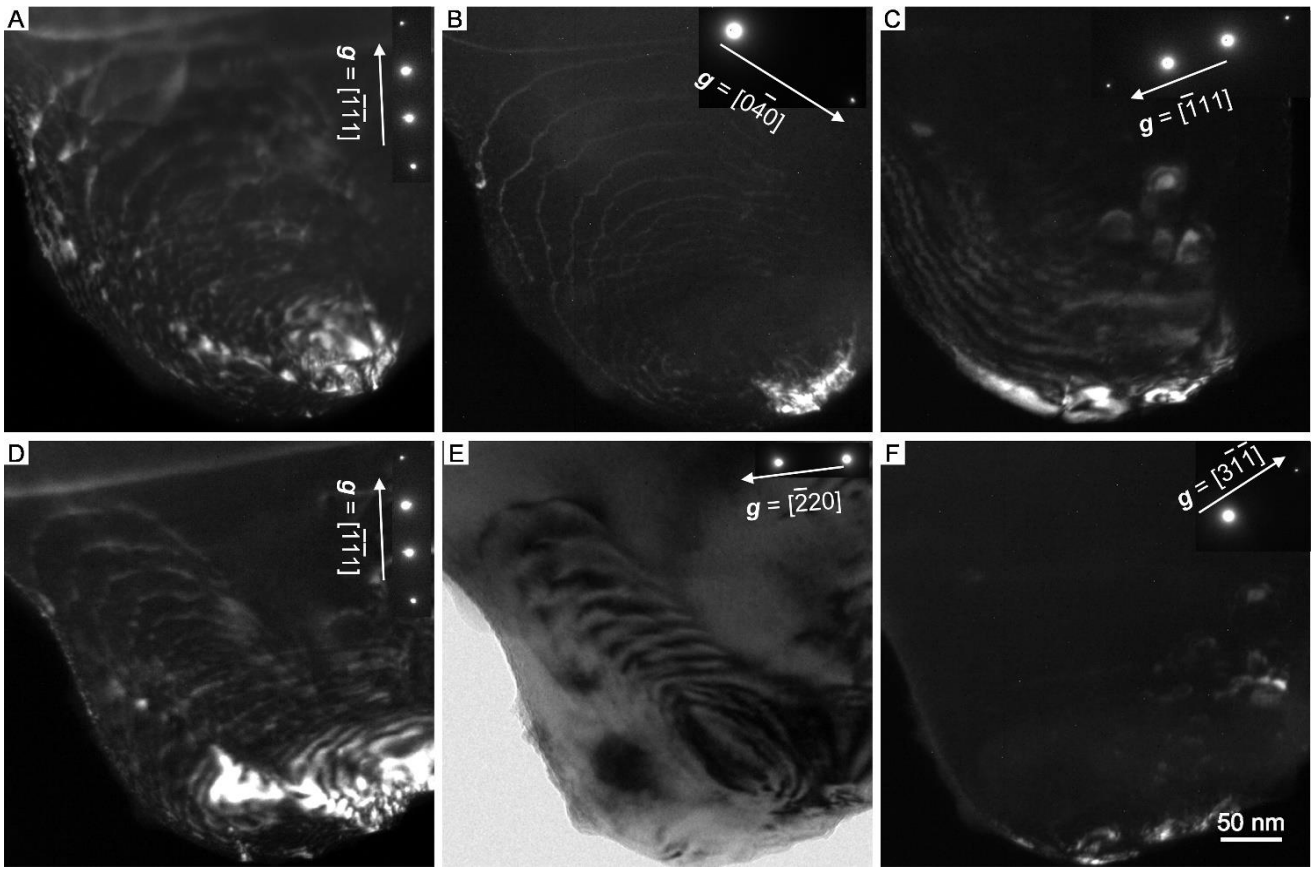

**Figure S4. *g·b* analyses of dislocations in a pillar.** The dislocations were taken under (A) ***g*** = [-1-11], (B) ***g*** = [0-40] and (C) ***g*** = [-111] two-beam conditions. In all three cases the sample is viewed along the [101] zone axis. (D) ***g*** = [-1-11], (E) ***g*** = [-220] and (F) ***g*** = [3-1-1] two-beam conditions. In these cases the sample is viewed along the [112] zone axis. Dislocations have strong contrast in (A), (B), (D) and (E), while showing extinction in (C) and (F). The Burgers vector was determined to be ***b*** =1/2 [01-1]. Insets show the corresponding SAED.

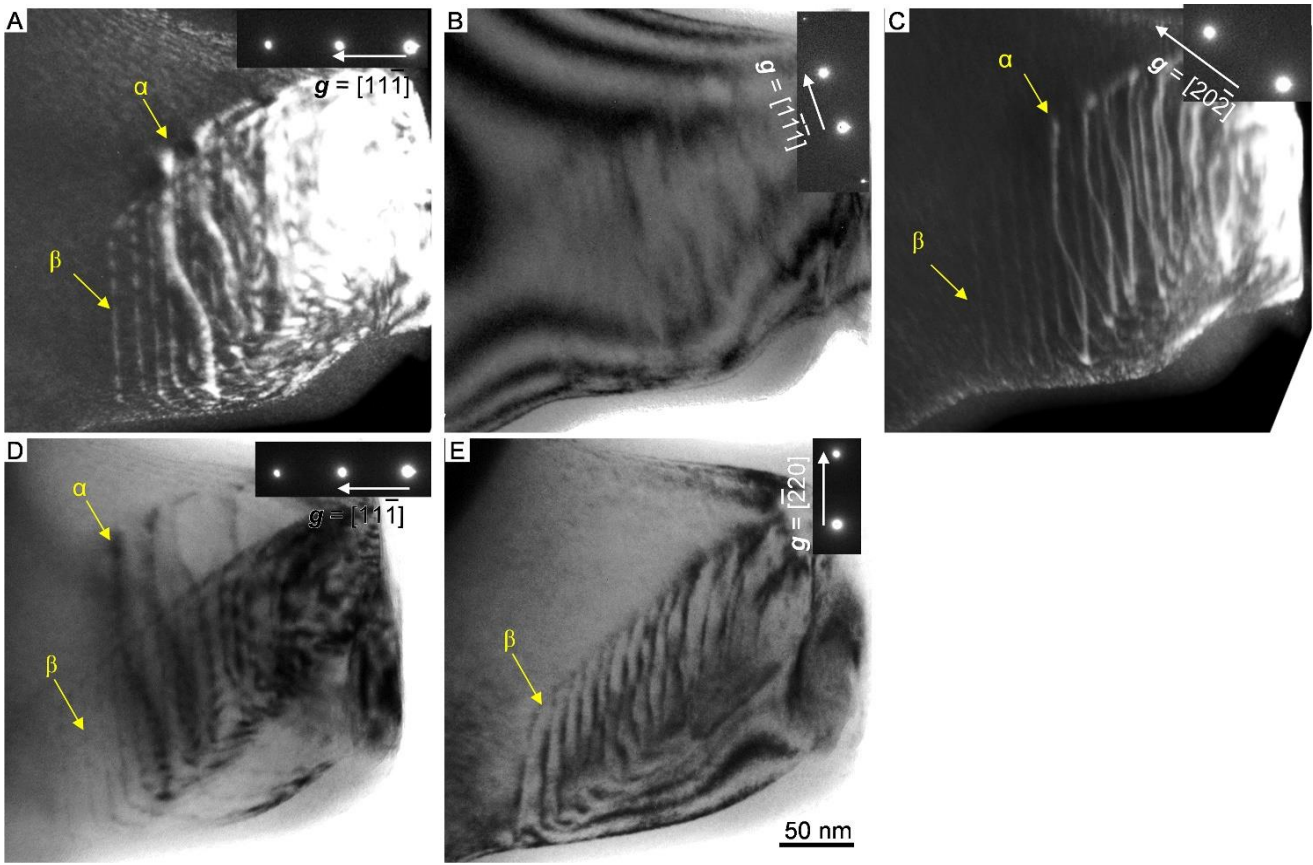

**Figure S5. *g·b* analyses of the dislocations in a diamond pillar.** The dislocations were taken under (A) *g* = [11-1], (B) *g* = [11-1] and (C) *g* = [20-2] two-beam conditions along the [101] zone axis, and (D) *g* = [11-1], (E) *g* = [-220] two-beam conditions along the [112] zone axis. Dislocation array α has strong contrast in (A), (C) and (D), while showing extinction in (B) and (E). Dislocation array β has strong contrast in (A), (C), (D) and (E), while showing extinction in (B). Notably, dislocations show residual contrast in (B), which is characteristic of edge-type dislocations. The Burgers vectors of dislocation array α and β are determined to be ***b*** = 1/2[110] and ***b*** = 1/2[01-1], respectively.

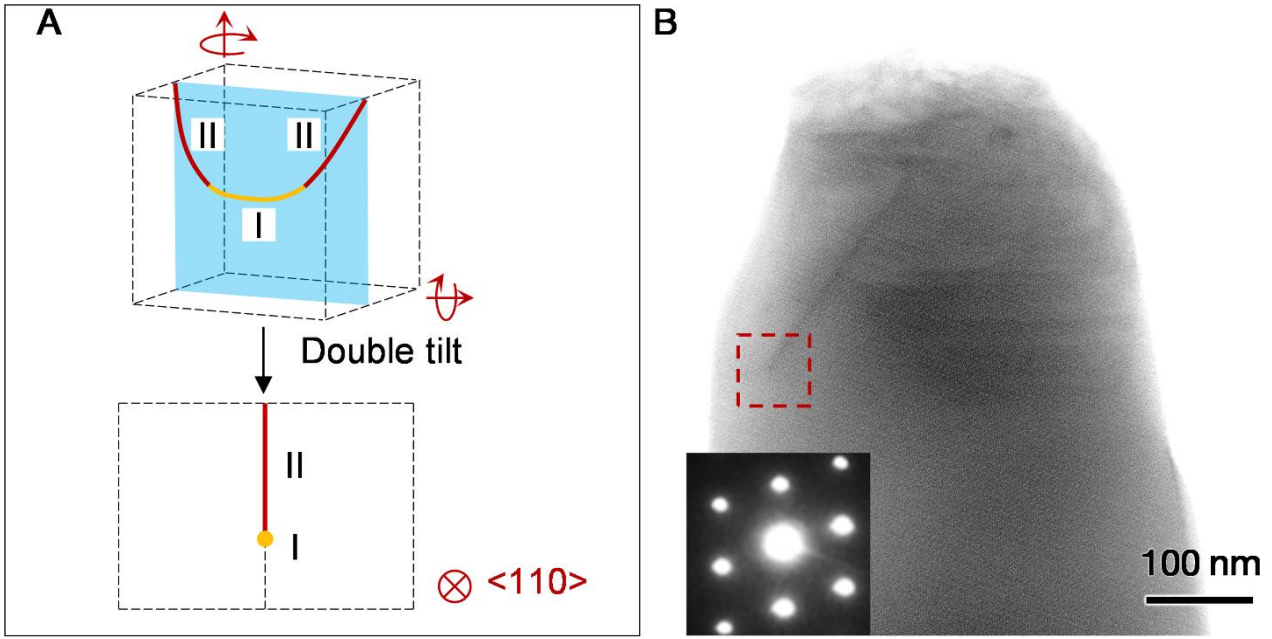

**Figure S6. The process to capture the dislocation core structure.** (A) In this schematic drawing, the sample is tilted to a certain <110> zone, and the slip plane is perpendicular to the page (the lower sketch). The yellow dislocation head segment (I) is viewed edge-on. (B) BF-STEM image of the sample after tilting shown in A. Inset shows the corresponding SEAD, the red square is where the core of dislocation head segment (I) located.

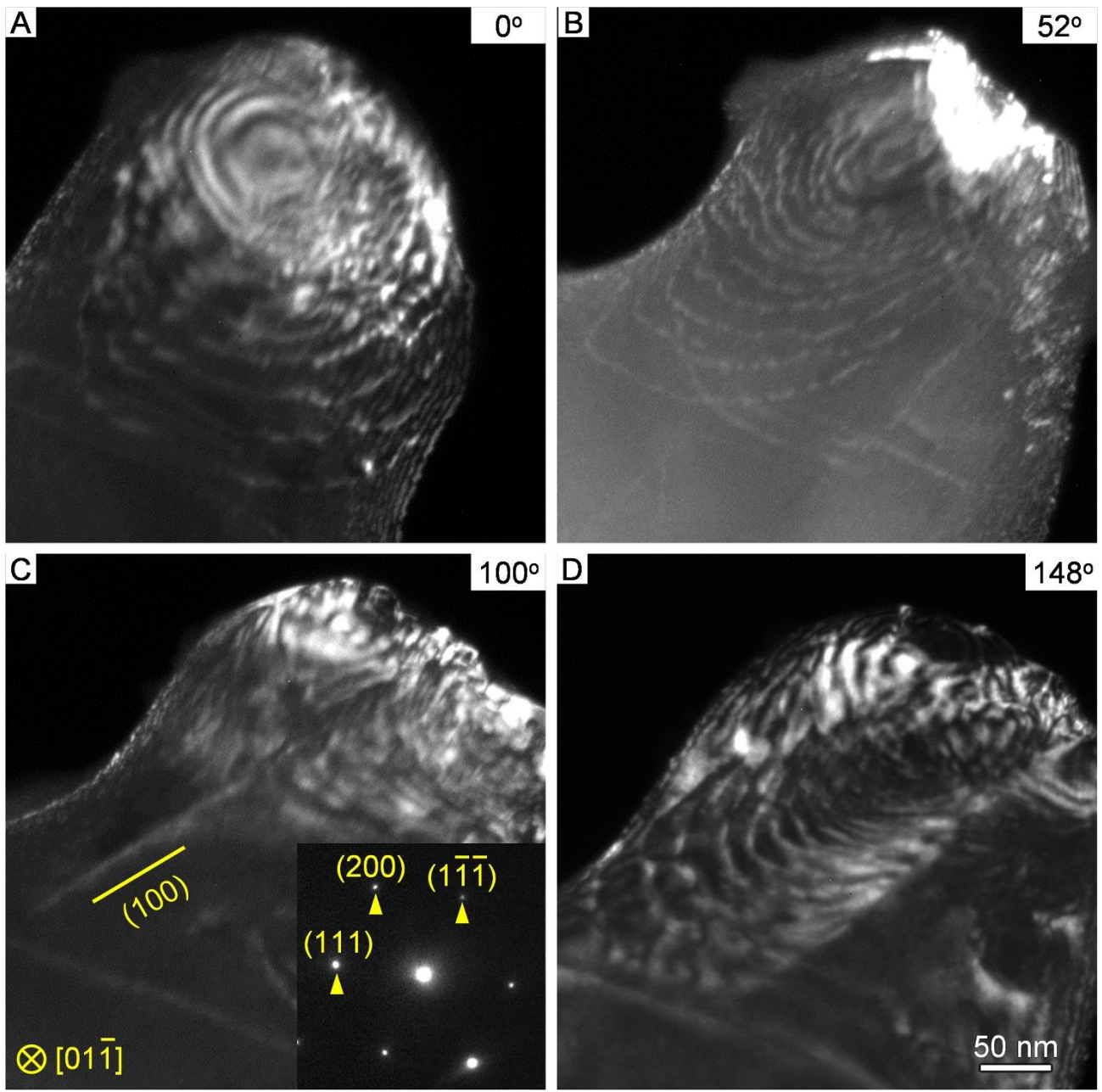

**Figure S7. Three-dimensional rotation of a diamond pillar.** (A-D) DF-TEM images of the compressed diamond pillar rotated at various angles (indicated). The rotation axis is near [1-1-1]. The slip plane is rotated perpendicular to the page in (C). The SAED (inset of (C)) shows that the dislocation slip plane is (100).

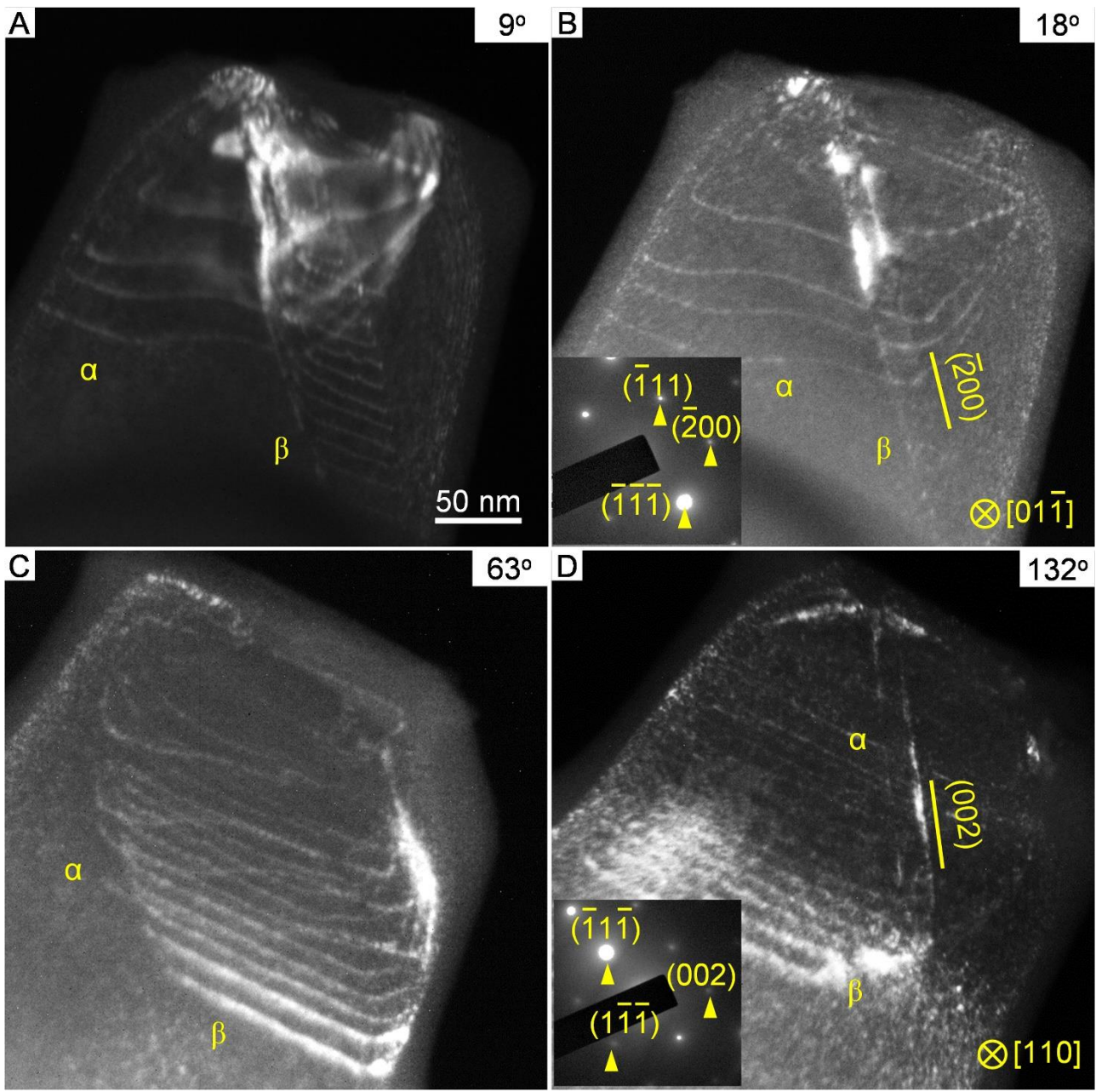

**Figure S8. Three-dimensional rotation of another pillar.** (A-D) DF-TEM images of the compressed diamond nanopillar rotated at various angles (indicated). The rotation axis is near [1-1-1]. The slip planes of dislocation array β and α are rotated to be perpendicular to the page in (B) and (C), respectively. SAED patterns (insets to (B) and (D)) show that dislocation arrays α and β slip on the (-100) and (001) planes, respectively.

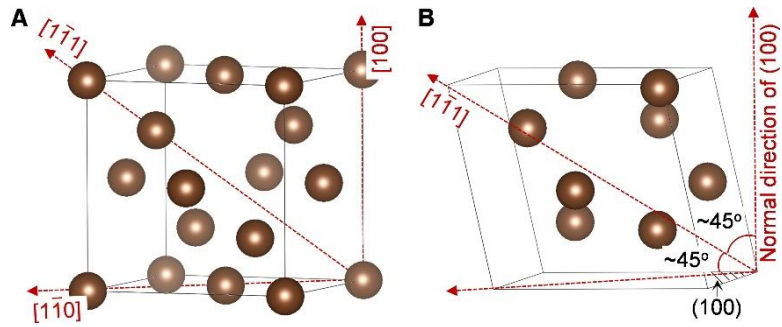

**Figure S9. The diamond lattice evolution under <111>-compression.** (A) The diamond lattice under zero load. (B) The severely deformed diamond lattice at the critical strain (~27%) resulted from our first principle calculations. The loading direction (i.e., <111>) has an angle of ~45° to both the slip plane (i.e., (100) plane) and slip direction (i.e., <110>) in such severely deformed structure.

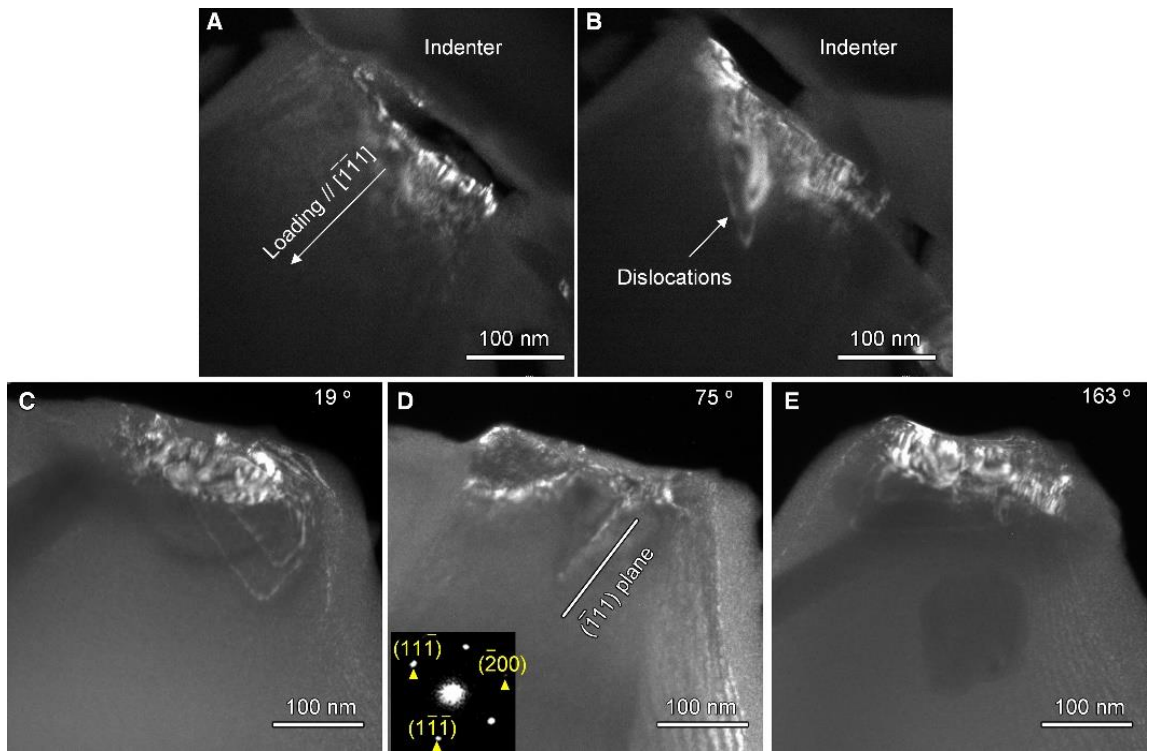

**Figure S10. Dislocation behaviors in <111>-compression silicon nanopillar.** (A and B) Dislocations are activated under load of <111> direction. (C-E) The sequences of produced dislocations in silicon through rotating the sample along <111> direction. Those results show the dislocations in <111>-compression silicon nanopillar slip in {111} plane.

**Table S1. The Schmid factors of each slip system on different loading directions**

| Loading direction | Slip system | Schmid factor |
|---|---|---|
| [111] | (100)[011] | 0.47 |

|  | (11-1)[011] | 0.27 |
| --- | --- | --- |
| [110] | (100)[011] | 0.35 |
|  | (11-1)[011] | 0.41 |
| [001] | (100)[011] | 0 |
|  | (1-11)[011] | 0.41 |